\documentclass[twocolumn,showpacs,showkeys,preprintnumbers,amsmath,amssymb,aps,floatfix]{revtex4-1}

\usepackage{amsfonts}
\usepackage{amssymb}
\usepackage{amsmath}
\usepackage{graphicx}
\usepackage{float}
\usepackage{color}

\begin{document}

\title{Actinide Topological Insulator Materials with Strong Interaction}

\author{Xiao Zhang$^{1}$,  Haijun Zhang$^{1}$, Claudia Felser$^{2}$, Shou-Cheng
Zhang$^{1}$}

\affiliation{
  $^1$Department of Physics, McCullough Building, Stanford University,
  Stanford, California 94305-404531\\
  $^2$Institut f\"ur Anorganische Chemie und Analytische
  Chemie, Johannes Gutenberg - Universtit\"{a}t,  55099 Mainz, Germany
}

\email{sczhang@stanford.edu.}

\begin{abstract}

Topological band insulators have recently been discovered in
spin-orbit coupled two and three dimensional systems. In this work,
we theoretically predict a class of topological Mott insulators
where interaction effects play a dominant role. In actinide
elements, simple rocksalt compounds formed by Pu and Am lie on the
boundary of metal to insulator transition. We show that interaction
drives a quantum phase transition to a topological Mott insulator
phase with a single Dirac cone on the surface.

\end{abstract}

\maketitle

Topological insulators are new states of quantum matter with a full
insulating gap in the bulk and gapless Dirac fermion states on the
surface\cite{Qi2009a, qi2011, hasan2010}. Spin-orbit coupling plays
a dominant role in topological band insulators HgTe,
Bi$_{1-x}$Sb$_{x}$ and Bi$_2$Te$_3$ class of materials. The basic
mechanism for all topological insulators discovered so far is the
band inversion driven by spin-orbit coupling\cite{bernevig2006d}.
Soon after the discovery of topological band insulators, the concept
of a topological Mott insulator has been proposed
theoretically\cite{raghu2008}, where interaction effects are
responsible for topological insulating behavior. The interplay
between electron interaction and spin-orbit coupling could enhance
the bulk insulating gap, which is important for device applications.
Magnetic and superconducting order driven by interactions could
co-exist with topological order, and give rise to exotic excitations
such as magnetic monopole\cite{qi2009b}, Majorana
fermion\cite{fu2008} and dynamic axion field\cite{Li2010}.

In this work we theoretically predict a new class of topological
Mott insulators in actinide compounds with electrons in the 5f
valence orbitals. Interaction and spin-orbit coupling energy scales
are comparable in these compounds, making them ideal material
systems to investigate the interplay between these two important
effects. Simple rocksalt compounds formed by Pu and Am lie on the
boundary of metal to insulator transition, where the 5f electrons
change from itinerant to localized characters\cite{moore2009}. Using
AmN as a example, we show that this rocksalt compound is metallic
without interactions, but an insulating gap is induced by
interactions. A band inversion between the 6d and the 5f drives the
system into a topological insulator phase with a single Dirac cone
on the surface. The predicted actinide topological insulators are
binary compounds with a simple rocksalt crystal structure and
reasonably large energy gaps, providing an ideal platform to
investigate interacting topological insulators. Standard topological
band insulators in the Bi$_2$Te$_3$ class of materials have bulk
carriers induced by disorder. In contrast, the actinide compounds
predicted in this work have stronger three dimensional ionic bonds.
As a result, they may have fewer bulk defects; in fact, earlier
photo-emission experiments indicate vanishing density of states at
the fermi level\cite{gouder2005}.

Americium and plutonium are actinide elements with electron
configurations of $5f^76d^07s^2$ and $5f^66d^07s^2$. Many of the Am
mono-pnictides AmX (X=N, P, As, Sb, and Bi) and Pu
mono-chalcogenides PuY (Y=Se and Te) are known to be nonmagnetic and
narrow-gap
semiconductors\cite{ghosh2005,suzuki2007,suzuki2009,gouder2005,ichas2001,
brooks1987}. At ambient conditions, they crystallize in the rocksalt
structure. The rocksalt structure has face-centered cubic (FCC)
symmetry with $Fm\bar{3}m$ space group, shown in Fig.~1A with AmN as
an example. It can be represented as two interpenetrating FCC
lattices with a two-atom basis. The first atom (Am) is located at
each lattice point, and the second atom (N) is located half way
between lattice points along the FCC unit cell edge. The bravais
lattice vectors are $a_1=(0, a/2, a/2)$, $a_2=(a/2, 0, a/2)$ and
$a_3=(a/2, a/2, 0)$. Its Brillouin zone (BZ) has three independent
time-reversal-invariant-momentum (TRIM) points, which are
$\Gamma$($000$), $X$($\pi\pi 0$) and $L$($\pi 00$) as shown in
Fig.~1B.
\begin{figure}[H]
\begin{center}
  \includegraphics[width=0.5\textwidth]{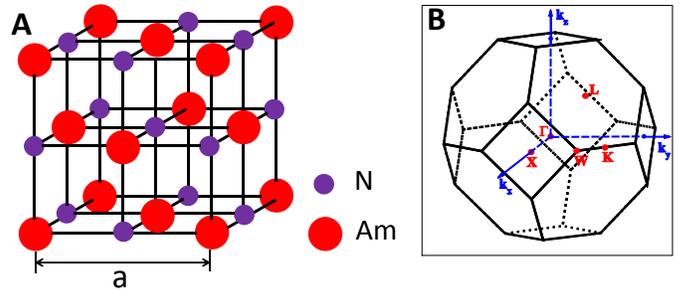}\\
\end{center}
  \caption{Crystal structure and brillouin zone. (A) Rocksalt crystal structure of AmN
   with $Fm\bar{3}m$ space group. Am and N separately form a face-centered cubic(fcc) sub-lattice,
   and these two sub-lattices are at a distance of $a/2$ away with each other along one cubic edge. (B) Brillouin zone
   for AmN of rocksalt structure. It has three independent time-reversal-invariant points are $\Gamma$($000$),
   $L$($\pi00$), and $X$($\pi\pi0$).}\label{crystal}
\end{figure}

In order to study the electronic structure of these materials, {\it
ab-initio} calculations are carried out in the framework of local
density functional approximation(LDA) supported by BSTATE package
\cite{fang2002} within the plane-wave pseudo-potential method. The
ultra-soft pseudo-potential\cite{vanderbilt1990} is employed for
localized orbitals, i.e., Am(Pu) 5f and N 2p orbitals. The
{\bf{k}}-point grid is taken as 16$\times$16$\times$16 and the
kinetic energy cutoff is fixed to 340eV for all the self-consistent
calculations. The spin-orbit coupling (SOC) interaction is taken
into account because both Am and Pu have very heavy nuclei. Due to
the localization of 5f electrons, LDA+U method\cite{Anisimov1993} is
used to deal with strong correlations in these compounds with
$U=2.5eV$ for Am \cite{ghosh2005} and $U=3.0eV$ for
Pu\cite{suzuki2009}. We use the experimental lattice constants in
our calculations\cite{Wastin1995}.

Because all the AmX and PuY have similar electronic structures, here
AmN is taken as an example. The band structures for AmN with $U=0eV$
and $U=2.5eV$ are shown in Fig.~2A-B separately. The strong
interaction U is responsible for the metal to insulator transition.
If we ignore the effect of U in the calculation, there are residual
5f states at the Fermi level ($E_F$), indicating metallic behavior
(Fig.~2A). Only by taking into count of $U=2.5eV$, those residual 5f
states are lifted away from $E_F$, and we can correctly predict AmN
to be a insulator (Fig.2~B), consistent with the experimental
finding\cite{gouder2005}. We will elaborate this point in the
following energy band analysis. First of all, we consider the Am-X
(Pu-Y) bonding. One $f$ and two $s$ electrons of Am bond with the
three $p$ electrons of pnictides, while two $s$ electrons of Pu bond
with the four $p$ electrons of chalcogenides. As a result, there are
six $f$ electrons left in the ground states for both AmX and PuY
before we consider the hybridization between $5f$ and $6d$ states in
Am or Pu. Since these strong bonding states are further away from
$E_F$ than the low energy Am-Am and Pu-Pu bonding states, we ignore
them in our band structure analysis in Fig.~3. In the following we
will just consider the Am-Am and Pu-Pu $d$ and $f$ low energy
bonding states. The $6d$ orbital first splits into $e_g$ and
$t_{2g}$ states due to the large crystal-field splitting (CFS) and
then splits again into $\Gamma_7^+$ and $\Gamma_8^+$ subbands due to
a small spin-orbit coupling (SOC) (Fig.~3)\cite{dresselhaus2008}.
The {\bf{f}} states split into $5f^{5/2}$ and $5f^{7/2}$ states with
angular momentum of $5/2$ and $7/2$, due to the SOC interaction. For
the $U=2.5eV$ case, the Coulomb U additionally shifts the $5f^{7/2}$
states away from $E_F$\cite{ghosh2005}, leading to the metal to
insulator transition (Fig.~2A-B). Therefore, only the $5f^{5/2}$
states are fully occupied with the six $f$ electrons with a
non-magnetic ground state, before hybridizing with $d$ orbital. CFS
further splits the $5f^{5/2}$ states into $\Gamma_8^-$ and
$\Gamma_7^-$ \cite{dzero2011}.

\begin{figure}[H]
\begin{center}
  \includegraphics[width=0.5\textwidth]{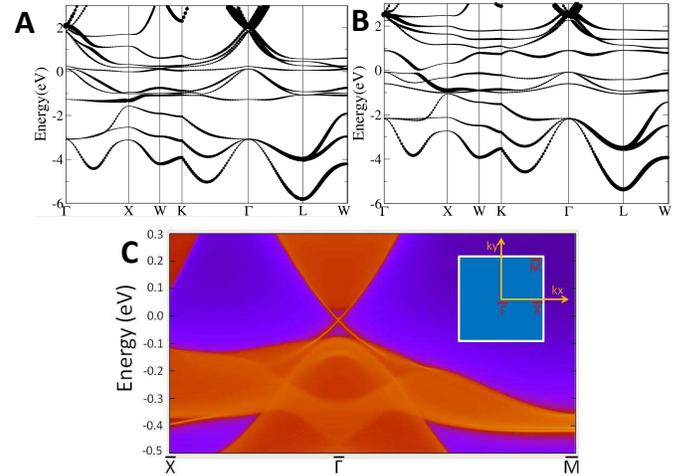}\\
\end{center}
  \caption{Band structure and topological surface states. Band structures for AmN
  at (A) $U=0eV$ and (B) $U=2.5eV$. The line width in these fat band structures corresponds
  to the projected weight of Am's d orbitals on Bloch states.
   The f states are divided into
two parts with angular momentum $5/2$ and $7/2$ due to the
spin-orbit-coupling interaction.
  The part with angular momentum 5/2 is occupied, and the other part with angular momentum $7/2$ is
  partially occupied with $U=0eV$. For $U=2.5eV$ case, these partially occupied f states of angular momentum $7/2$ move further toward high energy and become completely unoccupied. (C) Energy spectrum of the {\it ab-initio} calculation based TB
hamiltonian with the open boundary condition along '$+z$' direction.
One surface Dirac cone is clearly seen at $\bar{\Gamma}$ point as
red lines dispersing in the bulk gap on the AmN [001] surface. In
the inset, the projected two dimension BZ is shown.
  }\label{band}
\end{figure}
\begin{figure}[H]
\begin{center}
\includegraphics[width=0.5\textwidth]{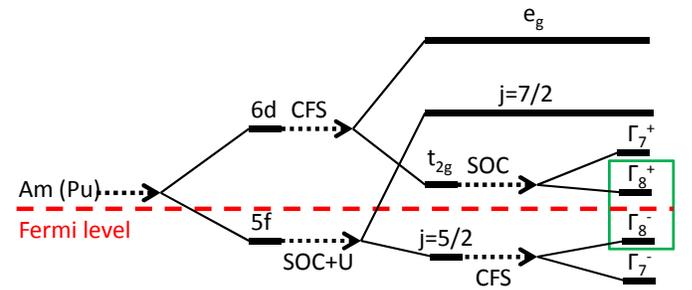}
\end{center}
\caption{Band sequence. Schematic diagram of the evolution from the
atomic d;f orbitals of Am or Pu into the conduction and valence
bands of AmX or PuY at the $\Gamma$ point. The black dashed arrow
represent the effect of turning on chemical bonding, CFS, SOC and U
(see text). The red dashed line represents the Fermi energy. The low
energy bands in the green box explains the topological physics.}
\label{figure3}
\end{figure}

\begin{table}
  \centering
  \begin{tabular}{|c|c|c|c|c|c|}
  \hline
     & $E_g$ & $\Gamma$ & 3$X$ & 4$L$ & Tot.  \\
  \hline
  AmN  & 0.100eV  & $-$ & $+$ & $-$ & $-$  \\
  AmP  & 0.085eV  & $-$ & $+$ & $-$ & $-$  \\
  AmAs & 0.080eV  & $-$ & $+$ & $-$ & $-$  \\
  AmSb & 0.055eV  & $-$ & $+$ & $-$ & $-$  \\
  AmBi & 0.040eV  & $-$ & $+$ & $-$ & $-$  \\
  PuSe & ~0.2eV      & $-$ & $+$ & $-$ & $-$  \\
  PuTe & ~0.178eV & $-$ & $+$ & $-$ & $-$  \\
  \hline
\end{tabular}
  \caption{Energy gap and parity eigenvalue. $E_g$ indicates the energy gap.
  $\Gamma$, $X$ and $L$ are the time-reversal-invariant(TRIM) points in Brillouin zone.
  '$-$' and '$+$' indicate the 'odd' and 'even' parity product for all the occupied bands.
  'Tot.' means the total parity product of all TRIM points. }\label{parity}
\end{table}

At $E_F$, the most important feature is the downward dispersion of
the $\Gamma_8^+$ suband of Am's d band, it crosses the $\Gamma_8^-$
subband of the $5f^{5/2}$ bands along $\Gamma-X$ direction, leading
to a band inversion near the $X$ point in the BZ (Fig.~3). A energy
gap is formed at the crossing point due to the f-d hybridization.
Because the d and f orbitals have 'even' and 'odd' parities at three
$X$ points, this inversion of bands with opposite parity signs
determines the non-trivial topology of these materials. This d-f
band inversion phenomenon is a generalization of the band inversion
between the s and p bands in HgTe quantum wells\cite{bernevig2006d}
and between p bands of opposite parities in Bi$_2$Te$_3$ class of
materials \cite{Zhang2009}. We calculate the parity at all TRIM
points\cite{fu2007b} (Table 1) and predict that AmN and all other
AmX materials in this family are non-trivial topological insulators
with their band gaps shown in Table 1. The LDA+U method cannot
obtain a energy gap for PuTe, which was measured experimentally to
be $0.2eV$ at ambient pressure and $0.4eV$ with pressure up to 5
GPa\cite{ichas2001,suzuki2009}. Suzuki and Oppeneer employ dynamical
mean-field theory (DMFT) method and successfully reproduce the
energy gap for PuSe and PuTe\cite{brooks1987,suzuki2009}. With the
same low energy physics as AmN, we can conclude that PuSe and PuTe
are non-trivial TIs.

According to the {\it ab-initio} calculation, the two low energy
$\Gamma_8^{\pm}$ bands can capture the salient topological features
of these materials. Therefore, we propose the following $8\times 8$
Hamiltonian:
\begin{equation}H=\left(
                    \begin{array}{cc}
                      H_d & H_{df} \\
                      H_{df}^+ & H_f \\
                    \end{array}
                  \right)
\end{equation}
$H_d$ and $H_{f}$ are $4\times 4$ Hamiltonian for $\Gamma_8^+$ and
$\Gamma_8^-$ representations with an effective angular momentum
$j=3/2$. We first write down the continuum Hamiltonian near the
$\Gamma$ point of the BZ and then extend it to a tight binding (TB)
model for the full BZ. The effective Hamiltonian for
$\Gamma_8^{\pm}$ bands is the Luttinger model \cite{luttinger1956}:
\begin{align}H_{d(f)}=C_{d(f)}+A_{d(f)}k^2+B_{d(f)}(k\cdot J_{3/2})^2
\\+D_{d(f)}(k_x^2(J_{3/2}^x)^2+k_y^2(J_{3/2}^y)^2+k_z^2(J_{3/2}^z)^2)\nonumber\end{align}
$A_{d(f)}, B_{d(f)}, C_{d(f)}, D_{d(f)}$ are constants and $J_{3/2}$
are spin matrices for angular momentum $3/2$. The first three terms
have continuous rotational symmetry; the last term breaks continuous
rotational symmetry while preserving cubic symmetry, which is
responsible for the band inversion only along $\Gamma-X$ direction
rather than other directions in the BZ\cite{sp}. The off-diagonal
Hamiltonian $H_{df}$ connects $d$ and $f$ bands of opposite
parities, therefore, the first order term can be written as
$H_{df}=F (k\cdot J_{3/2})$ with continuous rotational symmetry, $F$
again is a parameter.

Because the $5f$ and $6d$ orbitals are localized, we consider only
the nearest neighborhood hopping of Am-Am and Pu-Pu atoms. Each Am
(Pu) atom has twelve nearest neighbors located at $\pm a_1$, $\pm
a_2$, $\pm a_3$, $\pm(a_2-a_1)$, $\pm(a_2-a_3)$ and $\pm(a_3-a_1)$.
So we can extend the above continuum model to the TB lattice model
and reproduce the low energy bands of these materials well, with AmN
as an example\cite{sp}. Besides this single particle part, our
Hamiltonian also includes the on-site interaction parameter U for
electrons on the f orbital. One important effect the U interaction
is to renormalize the parameters of the single particle Hamiltonian,
and to induce a metal-insulator transition. On the insulator side,
we can effectively fit our model with renormalized parameters to the
first-principle calculations.

To observe the surface states, we construct a {\it ab-initio}
calculation for the semi-infinite AmN system with open boundary
condition in '$+z$' direction on the basis of maximally localized
Wannier functions (MLWFs)\cite{Marzari1997,Souza2001}. We can obtain
the surface Green's function of the semi-infinite system
iteratively. Finally the local density of states(LDOS) can be
calculated with the imaginary part of these Green's functions. From
these LDOS, we can obtain the dispersion of the surface states,
shown in Fig.~2C. A single Dirac cone at $\bar{\Gamma}$ clearly
shows up in the energy gap, which demonstrates the non-trivial $Z_2$
topology of the material in addition to the parity analysis. The
Fermi-velocity for the Dirac cone is $v_F \simeq 1.3\times 10^5m/s$,
a little smaller than Bi$_2$Se$_3$\cite{Zhang2009}. To confirm this
result, we also take open boundary condition in $z$ direction in the
analytical TB model (1)\cite{sp}, from which we have observed the
same single surface state Dirac cone at $\Gamma$ point of the BZ as
in the {\it ab-initio} calculation based TB hamiltonian.

\begin{acknowledgments}
We would like to thank Dr. Sri Raghu for useful discussions. This
work is supported by the ARO with grant number W911NF-09-1-0508.
\end{acknowledgments}

\end{document}